# Education dimensions of MGI: a question of multidisciplinarity, purpose, and equality

Eva M. Campo
Ronin Institute (eva.campo@ronininstitute.org)
University of Maryland at College Park (ecampo@umd.edu)

Executive Summary

This paper identifies the challenges associated with coordinating the development of new research methodologies and an accelerated pace of new discoveries in materials science with slower-evolving textbooks and curricula. The target audience is the undergraduate to post-graduate educational community, but implications for the K-12 pipeline are also considered.

This discussion is motivated by the Materials Genome Initiative (MGI), which aims to reduce the amount of time and resources needed to bring new materials to market. The MGI champions the development of materials through computational science (*in silico*) in close collaboration with *in experiment* materials design; advances are dependent on broad adoption of data science methodologies. The marriage between computational science and experimental techniques requires that we examine how we train future scientists and engineers. Can innovation in the classroom keep pace with the anticipated acceleration of research discoveries?

If the one end of the education pyramid (grad students/post docs) is excessively strained to deliver, for example, highly advanced mathematical/computational products as required by the MGI, are we at risk of producing an unsolvable divide in the educational structure? Further, are undergraduates being taught the right amount of mathematics, advanced science [e.g., quantum mechanics], and computation as well as data sciences? And is the STEM pipeline populated with students prepared to acquire the necessary skills and knowledge?

It is urgent to consider improving the pace of the innovation-to-textbook pathway. Currently, knowledge and technological development accumulate at the forefront of the scientific ecosystem, not necessarily propagating throughout the education chain, while more stringent demands are being placed on the workforce (i.e., use of sophisticated computational tools and knowledge of advanced scientific concepts). The question arises to whether a Golden Spike is likely to be achieved.

This paper proposes a lab-to-market classroom continuum with the theory-experiment-data intersection as the conceptual and content framework where the MGI is likely to unfold.

**Introduction**

Scientific initiatives have often proven to be efficient motivators in the realm of education to adequately train future workforce. Indeed, starting over a decade ago, nanotechnology has demonstrated that the seemingly disparate disciplines of physics and biology need interfacing. Novel education strategies can pave the way towards innovative scientific initiatives; for example, the Brain project can be traced to the earlier nano-bio-info-cognition paradigm championed by M. Rocco (Rocco, 2003). This multidisciplinarity momentum is now blurring the traditional boundary that the physical sciences have held where experimental and theoretical approaches have been kept apart. We find ourselves at a singular point in history where computational prowess and sophisticated experimentation merge to produce models and data with predictive quality (Jain, 2013), further optimizing lab-to-market pathways.

Some of the earliest discussions on the education implications of the MGI took place at the Spring 2013 Education Symposium hosted by the Materials Research Society. This symposium, inspired by the nanotechnology directives and the lab-to-market paradigm, led to the lab-to-classroom concept (Campo, 2013 and references therein). In particular, the nanotechnology initiative had identified the need to bring in domain scientists and end-users to expedite lab-to-market technology development through consumer acceptance (McNeil, 2007). Upon considering the classroom as an additional element in the framework, the lab-to-market-to-classroom (LMC) continuum has been introduced. This paradigm proposes a classroom model where practical, hands-on content of specific high interest to the students is not only delivered but also generated and end-consumer teste, creating a social ecosystem representative of society at large. Clearly, under a pedagogical description, the lab-to-classroom concept builds upon the notions of student-centered education and active and project-based learning (Strobel and van Barneveld, 2009). Project-based learning (PBL) is strongly student-driven and has been shown to improve learning attitudes, long-term retention, and skill development, often in interdisciplinary environments. An essential element, indispensable to both LMC and PBL, is that topics and systems under study are real problems and of specific interest to the students.

The MGI has identified the classroom as a component of the MGI framework to promote both technology development and market realization. From a pedagogical point of view, project-based learning and LMC share the pivotal concept of creating a motivation towards learning: why is it important to solve this problem and what tools do we need to solve it? Indeed, the MGI has strongly posed the question of lab to classroom- and in essence, the LMC is the conceptual framework of the MGI.

**The Advent of Data Science**

This discussion suggests that the pedagogical traditions where the MGI will flourish will likely be project-based learning within the K-12 to undergraduate curriculum. However, data science has been until now a missing element in that curriculum.

Data science encompasses data acquisition with comprehensive and accurate metadata, long-term preservation and curation, development of appropriate and efficient database schemas, and data mining and machine learning algorithms. There is an explosive growth in all of these areas in materials science, but it remains a challenge to educate the research community in best practices in the data life cycle. Indeed there is risk if traditional approaches of separating experiment and theory continue. In fact, the theory-experiment-data combination forms the content framework of the MGI, as seen in Figure 1, and in this context, the need to affect curriculum as well as availability of data science centers of expertise within experimental and theoretical user facilities clearly arises. Data science practices should be included as part of the curriculum in the near future taught alongside to standard error calculation. For the very large data sets that will be increasingly common and that require sophisticated data mining and machine learning algorithms, data science departments and professional data stewards are likely to become a necessary addition at national laboratories and research universities.

The "Towards a Lab-to-Classroom" MRS Education Symposium (Spring 2013) focused on the question of curriculum selection (White, 2013). The notion of a paradigm change has been proposed as a response to the growing concern that knowledge and technological development accumulate at the forefront of the scientific ecosystem, not necessarily propagating throughout the education chain, while more stringent demands are being placed on the workforce (i.e., use of sophisticated computational tools and knowledge of advanced scientific concepts). The question arises as to whether the ends will successfully meet in the middle.

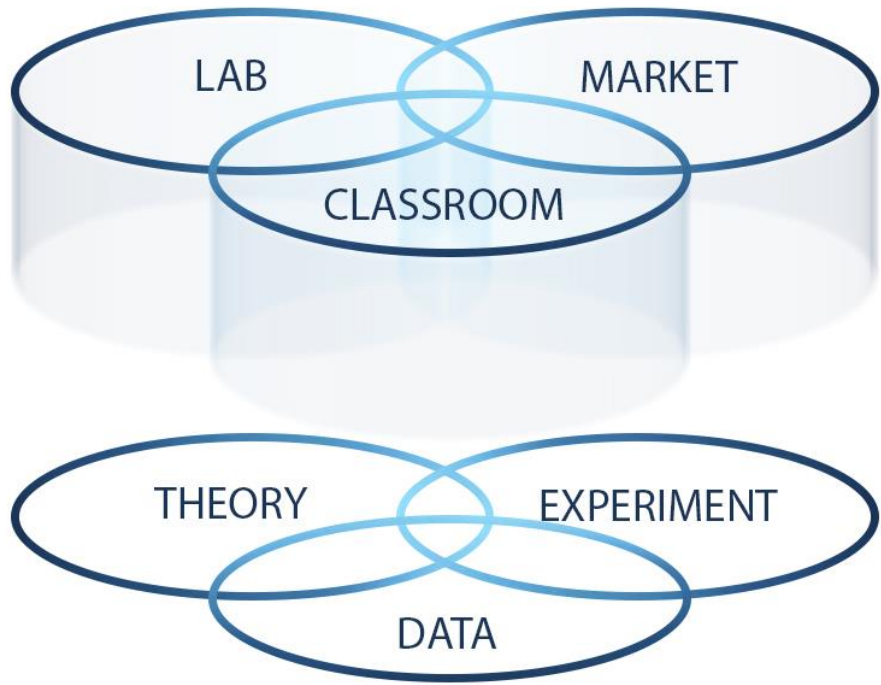

**FIGURE 1:**
**Introducing the Theory-Experiment-Data intersection and the Lab-to-Market-to-Classroom continuum:**
**Content and conceptual framework: the intersection zone at LMC is the conceptual framework of the MGI**

## Implications to the Educational Structure

As noted above, the very nature of the LMC and MGI seem to favor active learning by way of approaches such as project-based learning. LMC, and PBL suggest the need for whole school reform; the notion that a sustainable education paradigm, where new knowledge can be incorporated, may require decentralized education policy presents its own unique challenges (Cross, 2004). The identified LMC continuum could be the missing link to nurture sustainable scientific, technological, and curricular development. In this context, we see a new purpose to the MGI, as it could become a unique catalyst towards sustainable school reform. Indeed, this revised paradigm, could offer further insight towards prevention of static scientific curricula, and any foreseeable "tech-crisis".

These observations prompt the need for an urgent discussion to elucidate specific implications within two segments in the education ecosystem: K-12 and undergraduate/graduate. Our initial analysis suggests that modification of the conceptual framework is needed at the K-12 segment and a modification of the content framework is needed at the graduate/undergraduate segment.

**At the K-12 segment**, we propose that the LMC, an approach that follows a PBL paradigm, could be a viable conceptual framework that enables competence in the twenty-first century. As a proof of concept in this realm, the first LMC program has been designed for the dissemination of refreshable, photo-actuatable tactile displays to the visually impaired, serving both lab-to-market and lab-to-classroom initiatives (Campo, 2013). Indeed, by way of programmatic design through the logic model, this overlap was first predicted amongst classroom, market, and laboratory. Generally, this overlap will nucleate when a technology in developmental phase is deployed in a classroom with high affinity to such technology, in good agreement with lessons learned from PBL (Strobel and van Barneveld, 2009). In this scheme, students are stakeholders who help decide both content and applications to be included in the developing curriculum, and provide technology feedback, effectively leading to increased consumer acceptance. Notwithstanding the high value promised by PBL approaches, the need of whole school reform under this paradigm (Cross, 2004), certainly strains PBL, and hence, LMC implementation within the K-12 segment.

**At undergraduate, and especially graduate segments,** the adoption of students trained under LMC through K-12, will favor multidisciplinary learning as well as learning attitudes; paving the way towards the combined theory/experimental approach needed in MGI. In addition, as discussed earlier, methodologies to handle large data sets originating from high throughput experiments, data curation and data base tabulation, as well as procedures towards sharing, need to be elucidated. Of particular importance is the incorporation of adequate machine learning strategies. Absence of effective machine learning strategies to convert data into information, and subsequent information handling could open the door to uninformed systems over-characterization, yielding large data sets that are not adequately sampled for information.

## Community Building: Scientific Societies and Public

Professional scientific societies have already been identified as ideal spaces where education and scientific initiatives are first deployed, promoting both the quick advancement of nascent initiatives themselves, as well as the adoption of these initiatives worldwide, often assisted by new technologies. Indeed, early findings on the MGI and high throughput experiments (Green, 2014) as well as tutorials (Buongiorno-Nardelli, Marzari, Prendergast, and Rickman, 2014) have already been disseminated through innovative electronic media delivery. These newly offered capabilities offer singular opportunities towards inclusive practices, encouraging MGI participation from institutions with either limited or incipient research efforts such as community colleges or newly founded universities. This avenue, the dissemination of knowledge and newly found capabilities through electronic media, is likely to be explored by funding agencies and educational bodies alike.

Finally, and as a matter of relative urgency, we need to encourage the dissemination of the MGI as a new paradigm amongst the public, in the spirit of cultivating inclusion and participation of all stakeholders towards full development of this visionary initiative.

In summary, the MGI is an advanced multidisciplinary initiative, with profound consequences in both content and conceptual frameworks, as shown in Figure 1. It is at the intersection of data-experiment-theory in the content framework and on the intersection of lab-market-classroom in the conceptual framework, that the MGI will fruitfully unfold.